\documentclass[aps,prl,preprint]{revtex4}
\usepackage{color}
\usepackage[utf8]{inputenc}
\begin{document}

\title{Phase Space Formulation of Quantum Mechanics as an Hidden Variables Theory}

\author{ M. Revzen}
\affiliation{ Physics Department, Technion - Israel Institute of Technology,
Haifa 32000, Israel}


\begin{abstract}

An hidden variable (hv) theory is a theory that allows globally dispersion free  ensembles. We demonstrate that the Phase Space  formulation of Quantum Mechanics (QM) is an hv theory with the position q, and momentum p as the hv.\\

Comparing the Phase space  and Hilbert space formulations of QM we identify the assumption that led von Neumann  to the Hilbert space formulation of QM which, in turn,  precludes global dispersion free  ensembles within the theory. The assumption, dubbed I, is: "If a physical quantity $\mathbf{A}$ has an operator $\hat{A}$ then $f(\mathbf{A})$
 has the operator $f(\hat{A})$". This assumption does not hold within the Phase Space formulation of QM.\\

The hv interpretation of the Phase space formulation provides novel insight into the interrelation between dispersion and non commutativity of position and momentum (operators) within the Hilbert space formulation of QM and mitigates the criticism against von Neumann's no hidden variable theorem by, virtually, the consensus.

\end{abstract}


\maketitle

\section{Introduction}

The phase space formulation of quantum mechanics (QM) \cite{weyl,w,gron,moyal} (and numerous others \cite{zachos}) is a "logically
complete and self standing formulation" "independent of the conventional Hilbert space , or path integral formulations" \cite{zachos}.
The variables (observables) in theory are c-numbers functions of PS coordinates. I.e. it (the Phase Space theory) treats the position,q, and the momentum, p as defined simultaneously counter to the intuitive interpretation of QM based on the conventional Hilbert space formalism. This somewhat puzzling state was explicitly noted first , to our knowledge, by Leonhardt \cite{leonhardt} p.38.\\

A related idea is the a hidden variable (hv) theory pioneered by von Neumann  in his definitive monograph, "Mathematical Foundations of Quanatum Mechanics" \cite{vn}. Here an important notion is the
dispersion, D(A)  in the observed values of a physical quantity for identically prepared physical states:

\begin{equation}\label{dis}
D^2(\mathbf{A}) \;=\;[Exp(\mathbf{A}-<\mathbf{A}>)^2]\;=\;[Exp(\mathbf{A})^2]\;-\;[Exp(\mathbf{A})]^2\;= \;tr(\hat{\rho}\hat{A}^2)\;-\;
[tr(\hat{\rho}\hat{A}]^2.
\end{equation}

A globally dispersion free ensemble, $\hat{\rho}$, is one for which $[Exp(\mathbf{A}-<\mathbf{A}>)^2] \;\;\forall\;\hat{A}$ i.e. is independent of the observable.

von Neumann's definition \cite{vn} p.324 of an hv theory can be stated as follows. A theory is termed an hv theory if every observable's value can be expressed as a function of the coordinates of
a d-dimensional hv vector. Thus recalling that the value of an observable is an attribute of the ensemble under study (regardless whether
it preexisted in the ensemble prior to its measurement or not) an hidden variable (hv) theory has a dispersion free  value for every observable for every coordinate "point" of some vector space. I.e. in an hv theory
 an observable, $\mathbf{A}$, has a value $A = A(\mathbf{\lambda})$ for
$\mathbf{\lambda} =\lambda_1,\lambda_2,....\lambda_d$ and such dependences holds $\forall \mathbf{A}$  the  $\mathbf{\lambda}$ are the hv.
Implicit in this definition is that a coordinate (i.e. a "point" = $\mathbf{\lambda}$) of the hv space specifies a (hypothetical) state.(We shall return to this later.)
Thence, since within the Phase space formulation of QM every observable $\mathbf{A}$ has a dispersion free value for every phase space point, viz A(q,p), Phase space formulation is an hv theory with q,p as the hv.\\
An alternative definition is \cite{vn} p.326 is: an hv theory is a theory that allows a (globally) dispersion free ensemble i.e. a dispresion free "state" for all observables.\\
A brief outline of the Phase Space  formulation of QM is presented in Section II. Noting, as stated above, that within Phase Space  theory every
observable is given in terms of q,p thence via the first von Neumann's definition it (the Phase Space formulation) qualifies as an hv theory we prove, in Section III, that the theory allows a global dispersion free ensemble confirming thereby the theory as an hv theory according to VN's second definition. Indeed the proof underscores
the contention \cite{vn} p.324 that the two definitions are equivalent.\\

VN's study of "the entire structure of QM " p.295 led him to a formulation of the theory wherein the Hilbert space  formalism forms an integral part. Thence his theorem that the theory could not be embedded within a hv theory is a demonstration that "there exist no (globally) dispersion free ensemble" without a break with the Hilbert space formalism. This, he estimated to be "very unlikely in the face of its startling" success p.327. The identification of the
assumption within von Neumann's study of the "structure of QM"  entailing disallowance of embedding the theory within an hv theory attained a controversial status between the perhaps present consensus \cite{bell, mermin,ms,haag}, that von Neumann's theory is too restrictive
and thence inadequate and the  affirmation of the theory by Bub and Dieks \cite{bub2,dieks}. The identification of the Phase Space formulation as an hv theory allows us, in Section IV, to identify the the modification in von Neumann's assumptions that allowed hv formulation by direct comparison of the two theories (the Hilbert space and Phase Space formulations). Thus we argue, in Section IV, that the theorem is a tribute to von Neumann's mathematical insight that a hv theory is allowed only by a break from the HS formalism. The latter (i.e. the Hilbert space) involvement being induced via the seemingly "obvious" requirement dubbed assumption I by von Neumann p.313 viz
  $\mathbf{A} \leftrightarrow \hat{A}$ then  $f(\mathbf{A}) \leftrightarrow f(\hat{A})$. This assumption is violated within the Phase space formulation of QM allowing it, thereby, to form an hv theory.\\

The hv interpretation for the Phase space formulation  is suggestive for novel interpretations of some of its expressions. Thus, e.g., in Section V we consider the following: a well defined value of an observable $\mathbf{A}$ for an ensemble is attained for a dispersion free ensemble. The value in this case clearly equals the average value <A>. Within the Hilbert space formulation such dispersion free ensemble are the projectors $|a'><a'|$ to an eigern function of the the observable,
$\mathbf{A}$. Thus the ensemble has the dispersion free value a', the eigen value of $\hat{A}$ that represents $\mathbf{A}$. Mapping this to Phased Space we obtain a
dispersive distribution in the hv (i.e. q,p) variables that forms such  dispersion free ensemble. A fresh view, given in Section V, for the relation between
dispersion and operators non commutativity and a  novel interpretation of Wigner functions as dispersive (quasi) distributions of the hv for dispersion free distributions formed by projectors, of Hilbert space states. The last section, Section VI, contains some concluding remarks. An appendix provides an explicit example for dispersion of in the value of an observable implicit in the Wigner function of a projector of an eigen function.\\

\section{Brief outline of Phase Space formulation of Quantum Mechanics (QM)}

Our account of the Phase space  formulation is confined spinless non relativistic particle.  Though the theory  is, as stated above, an
independent discipline it is advantageous for our analysis to consider the theory as the 1-1 mapping of, the perhaps more familiar,
Hilbert space formalism (of non commuting operators and states) to the Phase Space one of c-numbers functions of (c-numbers ) q,p ,conceived by Weyl (1927).
\cite{weyl}. A central feature  of Weyl scheme is the mapping of c-number q,p polynomials to fully symmetrized , termed Weyl ordered, polynomials of
 $\hat{q},\hat{p}$. E.g. the Phase space c-number $q^2p$ is mapped to the Hilbert space Weyl ordered polynomials
 $\frac{1}{3}[\hat{q}^2\hat{p}+\hat{q}\hat{p}\hat{q}+\hat{p}\hat{q}^2]$.\\

The mapping violate von Neumann's assumption I p.313 in as much as while $\hat{A} \rightarrow a$ $f(\hat{A}) \rightarrow f'(a)\neq f(a)$, in general. E.g. \cite{case},

\begin{eqnarray} \label{case}
\hat{H}\;&=\;&\hat{p}^2/2m + m\omega^2\hat{q}^2/2 \rightarrow \tilde{H}=p^2/2m + m\omega^2 q^2/2, \nonumber \\
&&while \nonumber \\
\hat{H}^2\;& \rightarrow\;& \widetilde{H^2} = (p^2/2m)^2 + (m\omega^2 q^2/2)^2 + {\frac{\omega^2}{4}}(\widetilde{\hat{q}^2\hat{p}^2}+
\widetilde{\hat{p}^2\hat{q}^2})\nonumber \\
  &&\;=\;H^2 - \omega^2/4\;\neq\;(\widetilde{H})^2\;=\;H^2.
\end{eqnarray}

  The explicit formula for the "quantization mapping", i.e. $Phase space\rightarrow Hilbert space$, is given by

\begin{equation}\label{quant}
\hat{A}\;=\;A(\hat{q},\hat{p})\;=\;\frac{1}{2\pi}\int dqdp A(q,p)\mathbf{\Delta}(q,p),
\end{equation}

with

\begin{eqnarray}\label{DD}
\mathbf{\Delta}(q,p)&=&\frac{1}{2\pi}\int d\xi d\eta e^{i[\xi(\hat{q}-q)+\eta(\hat{p}-p)]};\;\;\;\; [\hat{q},\hat{p}]_{-} = i,\nonumber \\
tr \mathbf{\Delta}(q,p)\mathbf{\Delta}(q',p')\;&=&\; 2\pi \delta(q-q')\delta(p-p').
\end{eqnarray}

The inverse transformation, Weyl "de-quantization" (i.e. $Hilbert space \rightarrow Phase space$) that maps an Hilbert space operator $\hat{A}$ to a c number Phase space function of q,p $\tilde{A}(q,p)$ , is  given by,

\begin{eqnarray}\label{tr}
A(q,p)\;&=&\;tr\hat{A}\mathbf{\Delta}(q,p),\nonumber \\
        &=&\;\int dy e^{-ipy} \langle q+y/2|\hat{A}|q-y/2\rangle.
\end{eqnarray}

The Wigner function, W(A;q,p) \cite{w,moyal}, for arbitrary operator $\hat{A}$ is $W(A;q,p)\;=\;\frac{1}{2\pi}\tilde{A}(q,p)$.

A noteworthy feature of this formula is that it assigns simultaneously for all observables and without dispersion a value in terms of (values of)
 q,p.  Another feature of interest of this mapping is ,as noted above, the implied violation von Neumann's assumption I, viz given that an observable $\mathbf{A}$ represented by $A(\hat{q},\hat{p})$
within Hilbert space is mapped to $\tilde{A}(q,p)$ within Phase space, $f(\mathbf{A})$ represented by $f(A(\hat{q},\hat{p})$ within Hilbert space is  not mapped , in general
to $f(\tilde{A}(q,p))$ within Phase space. We wish to emphasize that this violation of is reflection onto the Phase space theory of the non commutativity of
of $\hat{q}, \hat{p}$ within the the Hilbert space formalism.\\

Of special interest in this connection is the mapping of a product $\hat{A}\hat{B}$ which turns out to be  the so called star product \cite{gron,moyal,zachos} of the mapped constituents:

\begin{equation}
\hat{A}\hat{B}\rightarrow A(q,p)*B(q,p), \;\;\;* \equiv e^{(i/2[\overleftarrow{\partial_q}\overrightarrow{\partial_p}\;-\; \overleftarrow{\partial_p}\overrightarrow{\partial_q}])}.
\end{equation}

Thence, for B=A, $\hat{A}\rightarrow A(q,p)$ while $\hat{A}^2\rightarrow A(q,p)*A(q,p) \neq A(q,p)^2$ as illustrated above for
A(q,p)=H(q,p) the harmonic oscillators energy.\\

It will be shown in the next section that it is the  violation of von Neumann's assumption I that epitomizes the departure of Phase space from the Hilbert space formalism and allows a hv account of QM.\\

\section{A Hidden Variables (hv) Theory:  Quantum Mechanics in Phase Space }

Quite generally a theory that allows globally dispersion free ensemble is dubbed \cite{vn} p.326 \cite{bub1}
 an hv theory. An
alternative, perhaps more intuitively appealing definition, \cite{vn} p. 324, for an hv theory is that therein
all observables $\mathbf{A}$ are given in terms of coordinates of some d- dimensional coordinate space
$\mathbf{\lambda} \;=\; \lambda_1,\lambda_2,...\lambda_d$, viz $A\;=\;A(\mathbf{\lambda})$. These two
(definitions) are equivalent as we show below and imply that PS is an hv theory with q,p as the hv.\\

The Weyl de quantization transformation, Eq.(\ref{tr}), is wherein representatives of (all) observables $\mathbf{A}$ in Hilbert space are mapped onto q,p space
by assigning a value for each at every q,p "point". We show below how a globally dispersion free ensemble can be defined for such set up i.e. wherein every
observable has q,p dependent value, by considering it as a particular case of a
 statistical layout.\\

Within Phase space a c- number function A(q,p) is assigned  to every observable $\mathbf{A}$,
 The assigned value, A(q,p),
is the Weyl transform of the Hilbert space operator corresponding to $\mathbf{A}$, cf. Eq.(\ref{tr}).\\

We now argue that a theoretical account for physical system wherein the value every observable $\mathbf{A}$ is given in terms of Phase space coordinates q,p allows  the definition of  globally dispersion free ensemble.\\

Consider discretized PS coordinates q,p. (A discretization scheme is outlined on p.310 of von Neumann's monograph \cite{vn}.)
Let the disceretized coordinates q,p form a probability space. I.e. an "experimental" outcome is a "point" q,p with the probability
P(q,p) that reflect the distribution (state) of the measured ensemble. The probabilities assigned to the measured ensemble has the following meaning. Consider an ensemble made of S'>>1 members each being a card marked with an (outcome) q,p. The cards are thoroughly mixed. An "experimental" run consists of S, S'>>S>>1, random "pickings" among the S' cards. If S(q,p) cards among the S picked are marked with q,p the probability of an outcome q,p is S(q,p)/S in the $S,S' \rightarrow \infty$ limit. A DF ensemble would be one with S(q,p)=S for some q,p. In such case the
the outcome of every experiment will equal its average value i.e. it is DF ensemble. For a physical theory wherein every observable
$\mathbf{A}$ has a value for each "point" q,p, the outcome q,p i.e. the assignment of the value q,p to the measured ensemble implies without dispersion a value to all observables. Thus for these hypothetical states (i.e. the phase space coordinate), termed hv states
a dispersion free ensemble is necessarily globally dispersion free ensemble. \\

Since physical systems within the Phase space formulation of QM have for every observable $\mathbf{A}$ a value A(q,p) for (every) phase space coordinate point,
 within the Phase space  theory a dispersion free ensemble is automatically globally dispersion free one. The value of the observable at q,p is the Weyl transform of the observable representative within
the HS formulation Eq.(\ref{tr}). This complete our demonstration that Phase space theory allows a globally dispersion free ensemble and thus qualifies as an hv theory.\\

Guided by Eq.(\ref{dis}) the equation for the dispersion free ensemble ,$\rho(q,p)$, for the value of observable $\mathbf{A}$ ,

\begin{equation}\label{disgdf}
D^2(A)=\int dqdp (A(q,p)-<A>)^2\rho(q,p)=\int dqdp A^2(q,p)\rho(q,p)- <A>^2\;=\;0.
\end{equation}

with $<A>$ is the the expectation value of $\mathbf{A}$.\\

The solution is :$\rho(q,p)\;=\;\delta(q-q')\delta(p-p')$ (for some q',p'), henceforth the dd distribution.\\

Within the dd distribution
(ensemble) the value of the observable $\mathbf{A}$ is given for q',p'  viz A(q',p'). Noting that every observable has its value
determined by specifying q',p', implies that the dd ensemble (distribution) is automatically globally dispersion free distribution. Thus the dispersion free ensemble for distributions of q,p is globally dispersion free distribution.\\

The dd ensemble is a globally dispersion free ensemble. Thus globally dispersion free ensemble  was shown to be allowed within the Phase space formalism. Hence the Phase space formulation is an hv theory. This
 conclusion is automatic upon using the alternative definition for
hv theory given above. We thus note that within the Phase space formualtion  every observable is a function of q,p - the hv.

\section{Dispersive ensembles}

By considering q,p as random variables it was demonstrated above that the values assigned to (all) observables via the
Weyl de quantizing transformation Eq.(\ref{tr}), is recovered as dispersion free expectation values of the observables for the dd ensemble:

\begin{equation} \label{rv}
<\mathbf{A}>_{dd}\;=\;\int dqdp A(q,p)\delta(q-q')\delta(p-p').\;=\;A(q',p').
\end{equation}

It was further noted that since the above result holds for all observables a dispersion free ensemble for q,p is necessarily a globally dispersion free
 ensemble. Recalling that von Neumann proved (p.321) that Hilbert sapce formalism cannot accommodate globally dispersion free ensembles. (I.e. Hilbert space formulation of QM is not an hv theory.) Noting further that the value of an observable is well defined only for a dispersion free ensemble, it is of interest to consider the (necessarily) dispersive distributions of q,p that correspond to Hilbert space dispersion free ensembles wherein  observables do have definite values. Thus
we seek the q,p distributions that yield definite values of quantum states,i.e. eigen function of Hilbert space self adjoint operators which are
dispersion free for measurements of the corresponding observables.\\

Returning to Eq.(\ref{dis}) the equation for a DF ensemble for measurement of $\mathbf{A}$ is

\begin{equation}\label{dishs}
D^2(A)=tr\rho(A) \hat{A}^2\;-\;(tr\rho(A) \hat{A})^2)^2 \;=\;0.
\end{equation}

It was proven by von Neumann (p.320) that the solution to this equation is $\rho(A)\;=\;|a'><a'|$ i.e. a projector onto an eif of $\hat{A}$
which therby form the dispersion free ensemble for measurements of $\mathbf{A}$. The value of $\mathbf{A}$ =a' equals it's expectation value for this state.\\

Note however that this ensemble, though dispersion free, is not globally dispersion free. Indeed as proven by von Neumann (p.320)  no globally dispersion free ensemble exist within the Hilbert space formulation - the theory is not an hv theory. The Weyl transform of $|a'><a'|$ is the Wigner function, W(a';q,p), \cite{w}, Eq.(\ref{tr}):

\begin{equation}\label{wf}
W(a';q,p)\;=\;\frac{1}{2\pi}\int dy e^{-ipy}<q+y/2|a'><a'|q-y/2>.
\end{equation}

 (Note: It is well known that the Wigner function is quasi distribution. We adopt the view that
for physical calculations we may treat it as proper distribution. To our knowledge this supposition has never led to any errors
\cite{leonhardt} perhaps suggesting the legitimacy of Feynman's negative probability \cite{feynman}.)\\

Weyl transforming $\ref{dishs}$ (cf. Eq.(\ref{dis})) gives the equation in terms of the q,p distribution. Its solution is
$ W(a';q,p)\;=\;\tilde{\hat{\rho}}(q,p)$:

\begin{equation}\label{disps}
\int dqdp A(q,p)*A(q,p)W(a';q,p)\;-\;(\int dqdp A(q,p)W(a';q,p))^2\;=\;0.
\end{equation}

The dispersion, D, within the Phase space formalism for the measurement of $\mathbf{A}$ of the ensemble (distribution) W(a';q,p), is

\begin{equation}\label{D}
D^2(A)]\;\equiv\;\int dqdp A(q,p)^2 W(a';q,p) \;-\;(\int dqdp A(q,p)W(a';q,p))^2)^2\;=\;D,
\end{equation}

utilizing  Eq.(\ref{disps}) gives  the value of D in terms of the the quantal entity $[A(q,p)A(q,p)\;-\;A(q,p)*A(q,p)]$, it is

\begin{equation}\label{D1}
D\;=\;\int dqdp ([A(q,p)^2\;-\;A(q,p)*A(q,p)]W(a';q,p)).
\end{equation}

As an example we return to having the measured observable be the energy of one dimensional harmonic oscillator in its ground state.
Thus for the representative (in Hilbert space) of $\mathbf{A}$ we have $\hat{H}=\hat{p}^2/2m + m\omega^2 \hat{q}^2/2$, its Weyl transform, i.e.
its phase space form, is $\widetilde{\hat{H}}\;=\;H\;= p^2/2m + m\omega^2q^/2.$ The Weyl transform of the Hilbert space ensemble
$|0><0|$ is, \cite{case},

\begin{equation}\label{W}
W"(0;q,p)\;=\;\int dy e^{-ipy}<q+y/2|0><0|q-y/2>\;=\;\frac{1}{\pi}e^{-(a^2p^2+q^2/(a^2))};\;\;a^2=1/(m\omega).
\end{equation}

This ensemble is manifestly dispersive

\begin{eqnarray}\label{qp}
<(q-<q>)^2>\;&=&\;\int dqdp q^2 W"(0;q,p)\;=\;1/(2m\omega);\;\;\;<q>=0. \nonumber \\
<(p-<p>)^2>\;&=&\;\int dqdp p^2 W"(0;q,p)\;=\;m\omega/2;\;\;\;\;<p>=0.\nonumber \\
<(q-<q>)^2> <(p-<p>)^2\;&=&\;1/4.
\end{eqnarray}

\begin{eqnarray}\label{hh}
\hat{H}^2\;&=&\;(\hat{p}^2/2m)^2 + (m\omega^2 \hat{q}^2/2)^2 + \omega^2(\hat{q}^2\hat{p}^2+\hat{p}^2\hat{q}^2) \rightarrow \nonumber \\
\widetilde{\hat{H}^2}\;&=&\;H(q,p)*H(q,p)\;= H^2 - \omega^2\; \neq \;(\tilde{\hat{H}})^2 \;=\;H^2.
\end{eqnarray}

Thence

\begin{equation}\label{D1}
D\;=\;\int dqdp [H(q,p)^2 -H(q,p)*H(q,p)]W"(0;q,p)\;=\;(\omega/2)^2.
\end{equation}

Thus whereas $|0><0|$ forms a dispersive free ensemble for $\hat{H}$,  it's phase space form, W"(0;q,p), is dispersive for H(q,p). A detailed example is given in the appendix.\\

Considerable interest exists within foundation of QM studies in the relation between operators non commutativity
and dispersion (e.g. uncertainty relation \cite{pier} and citations within). Such relation is implicit in
Eq.(\ref{D1}). Thus as outlined in section II tracking of the orderings of the operators $\hat{q}, \hat{p}$
or, perhaps on a deeper level, the violation of von Neumann's assumption I, results in the non vanishing of the difference $[H(q,p)^2-H(q,p)*H(q,p)]$. This difference determines, cf. $\ref{D1}$, the dispersion D.\\

It is the presence of A(q,p)*A(q,p) rather than $A(q,p)^2$ in $\ref{disps}$ reflecting, as noted above (cf,), the violation of a von Neumann's assumption I, which asserts the {\it dispersive} account of the hv formulae of the {\it dispersion free} ensemble of the Hilbert space formalism (cf.\ref{dishs}). \\

This understanding, it is suggested, underscores von Neumann's noteworthy insight in his no hidden variable proof that should have, perhaps, be better phrased
a proof that no globally dispersion free ensemble exists within the Hilbert space formalism.(Recall that to von Neumann in 1932 the only effective theory for QM was its formulation within  Hilbert space.) It is his  implied contention that it is the Hilbert space formalism that disallow hv underpinning. Since it is assumption I that forces the Hilbert space formalism it is it that need be modified to allow accommodating hv in the theory (QM).
 Taking advantage of the existence of an hv a theory, viz the Phase space formulation, we indeed note that it is this assumption that was abandoned to allow  a hv theory for QM. Thus von Neumann's limited applicability proof appear to us as his assignment of full weight of hv disallowance to the Hilbert space formalism.
 An insight that is corroborated within the available hv theory, the Phase space formulation of QM which indeed
 incorporates violation of assumption I and breaks away from the Hilbert space  formalism  (\cite{bub, dieks,ms}).\\

Recalling that a well defined value for an observable requires dispersive free ensemble. Though assured by von Neumann's no hidden variable theorem \cite{vn} p.320 that no globally dispersion free ensemble is allowed within the Hilbert space formulation of QM the theory does have well defined values for the observables: the dispersion free state projectors. These are, as noted above, necessarily dispersive within the Phase space formulation: they are the corresponding Wigner functions cf. Eq.(\ref{disps}).\\

This allows the following interpretation for the somewhat ad hoc expressions for the Wigner function \cite{royer}: they form the distribution (ensemble) of the hv (q,p) that has well defined expectation value. Thus for our harmonic oscillator example in its ground state:
W(0;q,p) is the distribution in the hv values for the harmonic oscillator state ($|0>$) whose energy is $\omega/2$\\.

\section{Summary and concluding remarks}

The phase space formulation of quantum mechanics (QM) is an independent formulation of the theory wherein observables and states
are c-numbers.
The theory  is reviewed and is shown to allow a global, i.e. obsevable independent, dispersion free  distribution (ensemble) thus qualifies as an hidden variable (hv) theory. The hv are q,p, the position q and momentum p.\\

Von Neumann's insight that an hv underpinning of the theory should require a break from an Hilbert space  formalism is briefly discussed and it is noted that, indeed, this break is confirmed within the Phase space  formulation. A globally dispersion free ensemble is not possible within Hilbert space formulation of QM.\\

Values of physical observables are well defined for dispersion free ensembles. These ensembles, within the Hilbert space formalism, are projectors on pure states. Transforming these pure state projectors to the Phase space formalism reveals the corresponding Wigner functions to be dispersive distributions of the hv (i.e. q,p). These dispersion abides by the uncertainty principle. Thus the dispersive hv (q,p), expressed via the Wigner functions forming their distribution, is shown to arise from the non commutativity of the Hilbert space operators $\hat{q},\hat{p}$. \\

Acknowledgement: numerous helpful comments and suggestions by Professor J. Bub and my coleagues Professors Pier Mello and Ady Mann are
gratefully acknowledged.

\section{Appendix: Dispersion of an Hilbert space eigenvalue}

Consider the example Eq.(\ref{W})-Eq.(\ref{hh}). The Wigner function Eq.(\ref{W}) with

$$E(q,p)\;=\;\frac{\omega}{2}(a^2p^2+q^2/a^2);\;\;\;a^2=1/(m\omega,$$

is given by

\begin{eqnarray}
W(0;q,p)dqdp\;&=&\;\frac{1}{\pi}\exp^{-(\frac{a^2p^2\;+\;q^2/a^2}{E_0})}dqdp,\;\;\;E_0=\omega/2. \nonumber \\
              &=&\;\frac{1}{2\pi E_0}\exp^{-\frac{E(q',p')}{E_0}}dq'dp',\;\;\;E(q',p')=q'^2+p'^2.\nonumber \\
              &=&\frac{1}{\pi E_0}\exp^{-E/E_0}dEd\theta,\;\;\;tan\theta=q'/p',\nonumber \\
W(0;E)dE\;\;\;&=&\;\exp^{-E/E_0}dE/E_0.
\end{eqnarray}

Thus,
\begin{equation}
<E>\;=\;E_0,\;\;\;\;\;<(E-<E>)^2>\;=\;E_0^2.
\end{equation}

Thence, cf. Eq.(\ref{D1}), the dispersion $D \;=\;<(E-<E>)^2>$.

\end{document}